# Active learning-driven uncertainty reduction for in-flight particle characteristics of atmospheric plasma spraying of silicon

Halar Memon[a], Eskil Gjerde[b], Alex Lynam[a], Amiya Chowdhury[a], Geert De Maere[b], Grazziela Figueredo[b*], and Tanvir Hussain[a*]

[a] Coatings and Surface Engineering, Faculty of Engineering, University of Nottingham, University Park, Nottingham NG7 2RD, UK;

[b] School of Computer Science, Faculty of Science, University of Nottingham, University Park, Nottingham NG7 2RD, UK

*Correspondence: G.Figueredo@nottingham.ac.uk; Tanvir.Hussain@nottingham.ac.uk

**Abstract**

In this study, the first-of-its-kind use of active learning (AL) framework in thermal spray is adapted to improve the prediction accuracy of the in-flight particle characteristics and uses Gaussian Process (GP) ML model as a surrogate that generalises a global solution without necessarily involving physical mechanisms. The AL framework via the Bayesian Optimisation was utilised to: (a) reduce the maximum uncertainty in the given database and (b) reduce local uncertainty around a contrived test point. The initial dataset consists of 26 atmospheric plasma spray (APS) parameters of silicon, aimed at ceramic matrix composites (CMCs) for the next generation of aerospace applications. The maximum uncertainty in the initial dataset was reduced by AL-driven identification of search spaces and conducting six guided spray trails in the identified search spaces. On average, a 52.9% improvement (error reduction) of RMSE and an $R^2$ increase of 8.5% were reported on the predicted in-flight particle velocities and temperatures after the AL-driven optimisation. Furthermore, the Bayesian Optimisation around a contrived test point to predict the best possible characteristics resulted in a three-fold increase in prediction accuracy as compared to the non-optimised prediction. These AL-guided experimental validations not only increase the informativeness of the limited dataset but is adaptable for other thermal spraying methods without necessarily involving physical mechanisms and underlying mechanisms. The use of AL-driven optimisation may drive the thermal spraying towards resource-efficiency and may serve as the first step towards fully digital thermal spraying environments.

**Keywords:**

Atmospheric Plasma Spray; Machine Learning; Bayesian Optimisation; Active learning; Thermal spray

## 1. Introduction

The performance and durability of critical components in high-value manufacturing sectors, such as those used in defence, aerospace, and automotive applications, are in a continuous drive to improve in the increasingly competitive global markets. Most Surface Engineering and Advanced Coatings (SEAC) technologies are under pressure to impart design iterations to converge a coating solution that can be rapidly incorporated in an agile end-user manufacturing environment with minimal optimisation. This is particularly challenging for complex processes, such as thermal spraying, that may involve nonlinear processing stages and interdependent functions. For a standard commercial powder feedstock, over 27+ process parameters needs to be optimised to achieve a coating with desirable microstructure and properties. Furthermore, there are more than 3000 companies in Europe with over a billion euros turnover that produce thermal spray coatings (Malamousi et al., 2022) and despite intense research and industrial efforts, only a few innovative/novel spray coating materials have earned commercial maturity in a reasonable timeframe. This is mainly due to the time constraints that come with the experimental nature and the efforts required to reduce variations and maximise performance for a given mode of operation. Machine Learning (ML) techniques are increasingly used to solve intricate and highly nonlinear processes without necessarily utilising the underlying physical mechanisms (Qiu et al., 2021; Tao et al., 2021; Xu et al., 2022) and from a machine-learning point of view, predicting the in-flight characteristics of thermal spray at given spray conditions for a desired coating can dramatically improve the experimentation throughput, thus, accelerating innovation and commercialisation.

The ML applications in thermal spraying have been growing since the last decade and have expanded to include the predictions of plasma sprayed in-flight particle temperature and velocities (Guessasma et al., 2003; Guessasma et al., 2004), particle diameters (Choudhury et al., 2011), deposition yield and porosity (Kanta et al., 2009), and the melting of the particles (Kanta et al., 2010, Tejero-Martin et al., 2019). The in-flight particle characteristics define the final deposited coating and the fluctuations in these characteristics may adversely affect the deposition (Bai et al., 2013; Lee et al., 2021). The studies have also expanded to include high-velocity oxy-fuel (HVOF) sprayed feedstock to examine the effect of the in-flight particle behaviour on the final coating (Liu et al., 2021a; Meimei et al., 2018). The use of Artificial Neural Networks (ANNs) in thermal spraying has been the centre point of most of the ML-driven work (Liu et al., 2021a; Liu et al., 2019; Paturi et al., 2021; Wang et al., 2021), due to its capability to differentiate into layers within the thermal spraying environment and the ability to make predictions based on a limited dataset. A few studies also compared the predictive accuracy of linear and non-linear ANNs models with other machine learning models, such as gradient boosting regression, decision tree regression, random forest regression, logistic regression, support vector machine, and *k*-nearest neighbour (Canales et al., 2020; Liu et al., 2021b). Yang et al. (Yang et al., 2022) trained a support vector machine model to help predict a set of optimised compositions of high-entropy alloys (number of features 10 and 371 datapoints extracted from literature), which resulted in a 24.8% higher hardness, when synthesised, and the hardness value was higher than the original dataset. The studies indicated that the selection of an ANN model was tailored for the specific application and the availability and fitting of data may impact the accuracy of predictions. The prediction accuracy of the ML models often depend on the diversity and quality of the dataset and an improvement in model predictability may be hampered when the data collection process is laborious, expensive, and resource-intensive, such as in the case of thermal spraying. Thus, there is a need for an ML approach that can be adopted to a wide range of thermal spray methods and be capable to guide towards more informative samples that reduce overall uncertainty in the given model.

A more promising approach for thermal spraying is developing an active learning (AL) framework that can guide the search for data points to improve the predictive performance of a surrogate ML model. In other words, an informed thermal spray data collection will increase the informativeness of the initial, limited dataset by identifying search spaces to sample, which will in turn, reduce the overall uncertainty in the surrogate model. In this paper, the first-of-its-kind use of the AL framework in thermal spray was utilised to reduce the maximum uncertainty and to improve the model predictability for desirable in-flight particle velocity and temperature characteristics. An initial dataset of the in-flight particle temperature and velocity of atmospheric plasma sprayed (APS) silicon particles over 26 different spray conditions were measured via an in-line thermal spray sensor. The dataset was then used to train two ML models (Random Forest (RF) and Gaussian Process (GP)) and the best model, in terms of their data fitting, prediction accuracy, and model stability, was employed as a surrogate (GP in this study) for the AL-based Bayesian optimisation framework. A total of 6 AL-guided thermal spray runs were carried out after the Bayesian optimisation. The guided data collection was used to reduce maximum uncertainty, or in other words, carry out sampling in unexplored regions within the dataset to improve the prediction accuracy.

## 2. Materials and Methods
### 2.1 Materials

Thermal spray grade silicon powder (Metco™ 4810, >99 wt. % Si) was supplied by Oerlikon Metco, USA. The irregular shapes of Si particles are shown in Figure 1a, with an inset magnified image of the same powder. Figure 1b shows the particle size distributions (PSD) of the powder, and the particles have $d_{10}$, $d_{50}$, and $d_{90}$ of 26.1 ± 2.4 µm, 42.1 ± 3.6 µm, and 84.2 ± 22.7 µm, respectively. The nominal PSD of -75 +15 µm advised by the manufacturer, measured via sieve analysis and laser diffraction (Microtac) analysis, correlates with the PSD measurements carried out in this study.

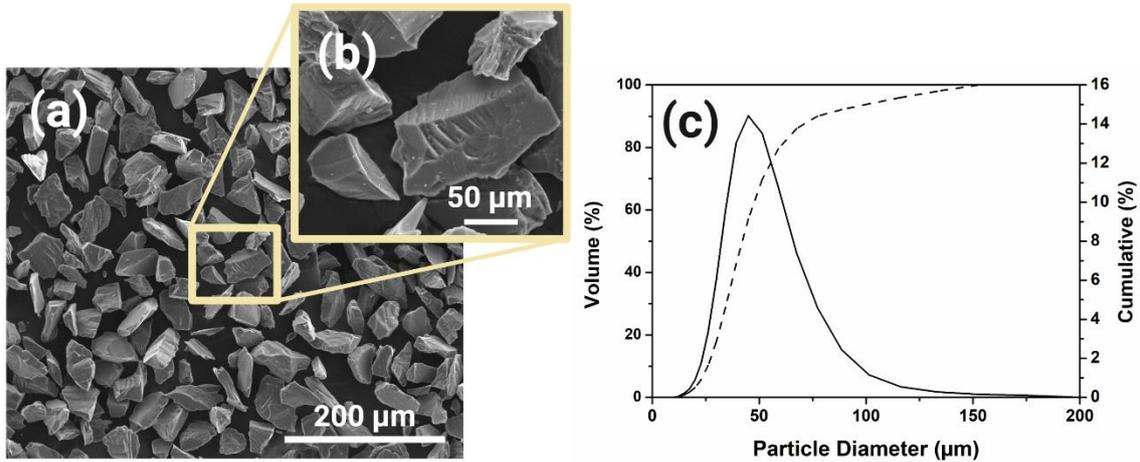

Figure 1: (a) The SE SEM Images and (b) particle size distribution of the silicon powder.

## 2.2 Atmospheric plasma spraying

All spraying trails were conducted using an SG-100 plasma spraying system supplied by Praxair Surface Technology, USA. The spray gun was fitted with a 03083-112 gas injector, a 02083-120 cathode, and a 02083-175 anode. The plasma torch used a mixture of argon and hydrogen gases to generate plasma plume, with Ar used as the primary gas and $H_2$ used as the secondary gas. A schematic of the plasma system is shown in Figure 2. More details on the atmospheric plasma spraying can be found in our previous studies (Lynam et al., 2022; Tejero-Martin et al., 2019).

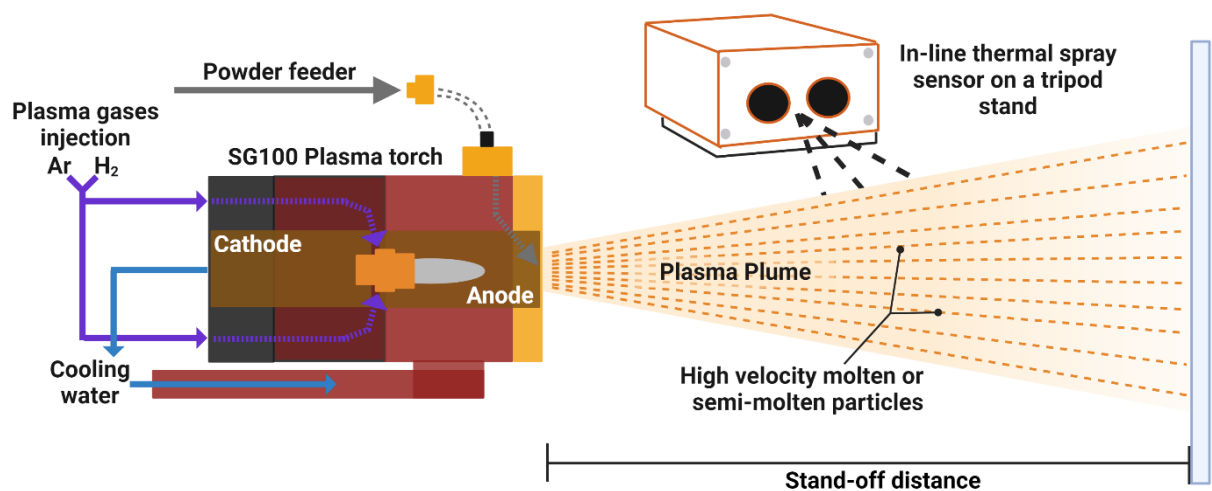

Figure 2: The schematics of SG-100 plasma spraying system with an in-line Accuraspray 4.0 thermal spray sensor.

## 2.3 In-flight particle characteristics measurement

The in-flight particle temperature and velocity measurements were carried out using an Accuraspray 4.0 thermal spray sensor, supplied by Tecnar Spray Sensors, Canada. The system is an in-line thermal spray sensor that measures the spray plume temperature (>1000 °C at 3% accuracy), velocity (5-1200 m/s at 2% accuracy), and stability (plume intensity at 2% accuracy and position at ±0.1mm accuracy) of the plume at a 400 mm field of view. The in-flight temperature and velocity are measured as an average of all the particles that pass through a large measurement volume of 750 mm$^3$ (3 mm x 25 mm x 10 mm) (Fauchais and Vardelle, 2010). The temperature was measured using a two-colour pyrometer, whereas the velocity was calculated when the particles travel between the two slits in the sensor, generating differentiating pulses. The response time was set to 1 s. The measurements presented in this study are a mean value of 60 measurements, which were obtained over a period of 60 seconds following a stabilised plume.

## 2.4 Machine learning models

All models were implemented in Python (Version 3.7.7) using the machine learning library scikitlearn (Version 0.23.2) and on a machine with a hardware configuration of Intel Core i9-9900K, CPU 3.60GHz, and 16GB of RAM. The hyperparameter tuning was carried out using scikit-learn's GridSearchCV, which performs an exhaustive search over a set of specified parameters. Two ML methods, Random Forest (RF), and Gaussian Process (GP), were studied to evaluate their applicability to the AL framework based on their data fitting, prediction accuracy, and model stability. Following evaluation, the GP model was selected as a surrogate model to implement the Bayesian optimisation framework. The AL using Bayesian optimisation iterations are detailed in a flow chart shown in Figure 3. At first, the initial dataset (spray trails 1-26) was used to train the GP model and the uncertainty areas within the initial data were identified to reduce maximum uncertainty. A guided uncertainty reduction was carried out using targeted further spray trails (spray trails 27-32), and the target processing parameter combinations were defined using the data search space. The new data was used to re-train the model, and the maximum uncertainty in the dataset was re-assessed.

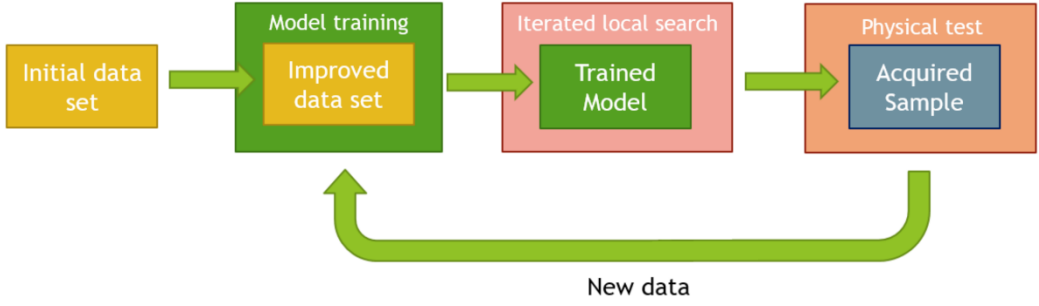

Figure 3: Schematic showing the AL-driven guided uncertainty reduction in the initial database.

The original spraying conditions and in-flight characteristics data were normalised between the values 0 and 1 to cast out any differential bias among the numerical features using the formula (Singh and Singh, 2020; Sola and Sevilla, 1997):

$$y_i' = \frac{y_i - y_{min}}{y_{max} - y_{min}} \tag{1}$$

where $y_i'$, $y_i$, $y_{max}$ and $y_{min}$ are the normalised data, original data, maximal data, and minimal data, respectively. To account for any outliers, the mean values were calculated using the modified Thompson tau test on the raw data (Anbarasi et al., 2011).

### 2.4.1 Random Forest

Random forest (RF) is a supervised learning method that uses ensemble learning for regression. The method combines multiple model predictions to make a more accurate prediction. In other words, the method works by constructing several decision trees that run parallelly with no interaction amongst them and the predictions out of these decision trees are governed in accordance with a random parameter, $k_{RF}$ (Biau, 2012). In the past, this has been shown to be an effective way to get substantial gains in classification and regression accuracy (Jin et al., 2020).

Random forest, a collection of randomised base regression trees, can be expressed as $\{r_n(\mathbf{X}_{RF}, \Theta_m, \mathcal{D}_n), m \geq 1\}$, where $n$ is the number of samples in the training sample $\mathcal{D}_n$ and $\Theta_1, \Theta_2, \ldots, \Theta_m$ are i.i.d. outputs of a randomised variable $\Theta$. The aggregated regression from the random trees can be estimated by (Biau, 2012):

$$\bar{r}_n(\mathbf{X}_{RF}, \mathcal{D}_n) = \mathbb{E}_\Theta[r_n(\mathbf{X}_{RF}, \Theta_m, \mathcal{D}_n)] \tag{4}$$

Where $\mathbb{E}_\Theta$ denotes the expectation with respect to the randomised parameter, which is conditionally attached with $\mathbf{X}_{RF}$ and the dataset $\mathcal{D}_n$. The optimal number of trees for this study was found to be 100, while the mean absolute error was used to determine the quality of a split.

### 2.4.2 Gaussian Process

The Gaussian Process (GP) Models are generic kernel-based supervised models designed to solve probabilistic and regression problems. In a multivariate case of the Gaussian process, there is a finite collection of random variables which have joint Gaussian (normal) distributions (Yom-Tov, 2004). The distribution over functions, $f(\mathbf{X}_{i\mathcal{GP}}) \sim \mathcal{GP}\left(\mu(\mathbf{X}_{\mathcal{GP}}), k\left(\mathbf{X}_{\mathcal{GP}}, \mathbf{X}_{\mathcal{GP}}'\right)\right)$, provided by the Gaussian process, is fully defined by its mean function $\mu(\mathbf{X}_{\mathcal{GP}})$ and the covariance is given by the kernel function $k\left(\mathbf{X}_{\mathcal{GP}}, \mathbf{X}_{\mathcal{GP}}'\right)$ (Guidetti et al., 2021). In a case of considerable variability in the available data, noise measurement, $\epsilon_{i\mathcal{GP}}$, within the distribution $\mathcal{N}(0, \sigma^2{}_n)$ is included in the final equation (Ebden, 2015):

$$y_{i\mathcal{GP}} = f(\mathbf{X}_{i\mathcal{GP}}) + \epsilon_{i\mathcal{GP}} \qquad (5)$$

Where $i$ is the $i$-th measurement corresponding to an input vector $\mathbf{X}_{i\mathcal{GP}}$. By including some known points $\mathbf{X}_{\mathcal{GP}}$ of the dataset and by identifying some unknown points $\mathbf{X}_{*\mathcal{GP}}$ for the estimation $f(\mathbf{X}_{*\mathcal{GP}})$, the model processes the probability distribution $p(f_*|\mathbf{X}_*, \mathbf{X}, f)$ with an assumption that the joint distribution $p(f_*|f)$ is jointly Gaussian, leaving the mean $\mu_*$ and covariance matrix $k(\mathbf{X}_*, \mathbf{X}_*')$ that define the final distribution $f_* \sim \mathcal{N}(\mu_*, \Sigma_*)$ (Berrar, 2019). By sampling from the final distribution, the desirable estimation can be obtained.

The RBF is also used as the kernel for the GP model. The kernel hyperparameters were optimised by maximising the log-marginal-likelihood during the fitting of the Gaussian process model. The level of noise in the available dataset was found to be 0.01. The prior mean is initialised to be the training data mean.

### 2.4.3 Stability of the models

The stability of the models was ascertained by the relative change percentage in the evaluation metrics between the training and testing accuracies (Fan et al., 2018). The relative change percentage $\delta_{i,j}$ between the two datasets, the evaluation metric $i$ and a model $j$, can be calculated using:

$$\delta_{i,j} = \left|\frac{\delta_{i,test} - \delta_{i,train}}{\delta_{i,train}}\right| x\ 100\% \qquad (6)$$

Where $\delta_{i,test}$ and $\delta_{i,train}$ are the relative change percentages of testing and training datasets, respectively.

### 2.4.4 Comparison of Model and Statistical Error Analysis

To evaluate and compare the accuracy and performance of the studied models for predicting in-flight particle characteristics, three common statistical metrics were used. They were root mean square error (RMSE), mean absolute error (MAE), and the coefficient of determination ($R^2$). The mathematical equations for the statistical metrics are shown below:

$$RMSE = \sqrt{\frac{\Sigma_{i=1}^{N}(y_i - y_i^*)^2}{N}} \quad (7)$$

$$MAE = \frac{1}{N}\Sigma_{i=1}^{N}|y_i - y_i^*| \quad (8)$$

$$R^2 = 1 - \frac{\Sigma_{i=1}^{N}(y_i - y_i^*)^2}{\Sigma_{i=1}^{N}(y_i - \bar{y})^2} \quad (9)$$

Where $y_i$, $y_i^*$, and $\bar{y}$ are the original values, predicted values, and the mean of original values, respectively. $N$ is the total number of predicted values. For RMSE and MAE, lower values indicate a better model performance, while for $R^2$, the closer it gets to 1, the better the regression line will fit the data.

### 2.4.5 *k*-Fold cross-validation

The *k*-Fold cross-validation is a statistical evaluation method used to ascertain the suitability of a machine learning model to predict the outcome of unseen data. In other words, the cross-validation estimates the capability of the model to generalise/make predictions on the data that was not used during the training of the model (Berrar, 2019). The validation works by shuffling and splitting the data into non-overlapping $k_{fold}$ number of smaller sub-sample folds, that are equal in size and the number is less than or equal to the number of elements in the dataset. This ensures that all of the available data (not just the explored regions) is used for standard train-test splits, and the model is trained on the new training set. The model is trained with the first subset as the test data and the rest of the subsets as the training data. The total error rate is calculated when the trained model is used to predict the test data, and the training iterations of the model continue above the *k*-fold value (which means the *k*-number of error rates). The total error rate is averaged from the *k*-number of error rates, essentially validating the generalising capability of the model. For the cross-validation, *k* was set to 15, which is close to what is commonly suggested in the literature (Marcot and Hanea, 2021).

### 2.5 Bayesian Optimisation

Bayesian optimisation is a sequential search framework (AL) that incorporates both exploration and exploitation of the available dataset. The framework consists of two main components. (1) A surrogate model is used as a proxy for the statistical/probabilistic modelling of the objective function, which is in-flight particle characteristics in this study. Between GP and RF regression, the GP model was used as the surrogate model to explore and estimate the maximum certainty in the database. (2) An acquisition Function, as a metric function, programmed to generalise a parameter that will consistently return the best optimal value from the optimised database. The acquisition function uses an exploration vs exploitation strategy to decide the optimal parameter. For example, while doing the hyperparameter space exploration, the framework searches for multiple optimal values in a given search area, and importance is given to the consistent optimal points (either high or low). Within these explore and exploit iterations, the surrogate model helps to get a simulated output of the function. By using the mean and variance for each point $\mathbf{X}_{BO}$, an acquisition function can compute the desirability of sampling at that location. More importantly, the goal is to find the maximum point using the minimum number of functional iterations. Given a function $f(\mathbf{X}_{BO})$ that estimates the coating quality in a form of a numerical value, the aim is to find the processing parameter combination $\widehat{\mathbf{X}}_{BO}$ that maximises the output of the function, over some domain $\mathcal{X}$ that has finite upper and lower bounds on each variable. The function is written as:

$$\widehat{\mathbf{X}}_{BO} = \mathrm{argmax}(f(\mathbf{X}_{BO})) \qquad (10)$$
$$\mathbf{X}_{BO} \in \mathcal{X}$$

The domain $\mathcal{X}$ consists of four variables in our problem, which are argon pressure, hydrogen pressure, arc current, and standoff distance. The additional variables in the database, particle temperatures and velocities, are dependent on the four controllable variables and are not included in the domain. The upper and lower bounds for each variable are detailed in Table 1. The upper and lower bounds were determined based on the plasma spray equipment limitations, and the Standoff distance (SOD) was based on the experimental limitation for the given thermal spray method.

Table 1: The upper and lower bounds of each variable for the Bayesian optimisation function.

| Variable | Lower bound | Upper bound |
| --- | --- | --- |
| Argon (psi) | 45 | 95 |
| Hydrogen (psi) | 25 | 100 |
| Current (A) | 25 | 650 |
| SOD (mm) | 50 | 200 |

In order to derive the desired acquisition function, i.e. striking a balance between exploration and exploitation, the metric known as Expected Improvement (EI) is used. Other potential metrics that improve the potential function may include the Probability of Improvement (PI); however, the metric only takes into account the likelihood that a point will return a result higher than the current maximum while ignoring the potential improvement. The EI weighs an expected improvement whenever a new data point is tried, and the EI can be calculated by the following formula:

$$EI(\mathbf{X}_{BO}) = \begin{cases} (\mu(f(\mathbf{X}_{BO})) - \max(f(\mathbf{X}_{BO}))\,\Phi(z) - \sigma(f(\mathbf{X}_{BO}))\emptyset(z), & if\,\sigma(f(\mathbf{X}_{BO})) > 0 \\ 0, & if\,\sigma(f(\mathbf{X}_{BO})) \leq 0 \end{cases} \quad (11)$$

$$z = \frac{\mu(f(\mathbf{X}_{BO})) - \max(f(\mathbf{X}_{BO}))}{\sigma(f(\mathbf{X}_{BO}))} \quad (12)$$

where $f(\mathbf{X}_{BO})$ is the surrogate model, $\mu(f(\mathbf{X}_{BO}))$ and $\sigma(f(\mathbf{X}_{BO}))$ are the prediction and uncertainty at a given location $\mathbf{X}_{BO}$, $\Phi(z)$ is the standard normal cumulative probability density, $\emptyset(z)$ the standard normal probability density, and $\max(f(\mathbf{X}_{BO}))$ is the maximum EI for the current set of samples used to train the surrogate model.

GP regression was used as the surrogate model because it performed the best compared to the RF model. To obtain a location in the data search space where a data point needs to be sampled, the Iterated Local Search (ILS) was used. The ILS is a stochastic global optimisation algorithm that works by repeating a local search algorithm on the modified versions of a good candidate solution. The search was randomly reset $N_{resets}$ times to find the global optimum out of the local optima. The local optima were found by repeatedly updating $S_*$ (random initial solution). The new candidate solutions were first generated by perturbing $S_*$, and the new candidate solution $S'$ was accepted as the new $S_*$ if the expected improvement of $S'$ is greater than the expected improvement of $S_*$. After no further improvements possible, the best local optima is then described as $S_{best}$.

## 3. Results

### 3.1 Initial experimental data collection

A total of 32 different spray trials were conducted in this study, and the spray conditions are specified in Table 2. The spray trails 1-26 were used as the initial database to process the ML models fitting and stability. The spray conditions were designed to reflect a wider range of processing parameters, i.e. different gas flows of primary and secondary gases, the overall current of the spray torch, and the SODs from 50 mm and up to 150 mm, and the conditions produced diversified in-flight characteristics to reflect a wide spectrum of particle temperatures and velocities. The spray trails 27-32 were conducted as part of the AL-driven Bayesian optimisation cycle.

During the initial spray trails, the in-flight particle temperatures ranged from 2370 °C up to 2883 °C and the inflight particle velocities from 116 m/s up to 242 m/s. An increase in the primary Argon and secondary hydrogen gases and the current of the torch resulted in an increase in particle temperatures and velocities. Furthermore, the effect of current on particle temperatures and velocities is more pronounced at larger SODs, or in other words, the effect of current on particle temperatures and velocities seems to diminish when the SODs were reduced to 50 mm. Increasing the SODs from 120 mm to 150 mm results in a 12-14% decrease in particle temperatures and a 2-4% decrease in particle velocities at 500 A current. At 600 A current, the reduction nearly halves to a 7-8% decrease in particle temperatures and a 1-2% decrease in particle velocities, as an increasing current of the torch increases the burning capacity of the plasma plume.

Table 2: Summarising the spray trials conducted in this study

| Spray Run | Ar (psi) | $H_2$ (psi) | Current (A) | SOD (mm) | Temperature* (°C) | Velocity* (m/s) |
|---|---|---|---|---|---|---|
| Initial database | | | | | | |
| 1 | 65 | 30 | 500 | 120 | 2588.6 | 183.6 |
| 2 | 85 | 30 | 500 | 120 | 2601.2 | 194.3 |
| 3 | 65 | 35 | 500 | 120 | 2604.2 | 186.8 |
| 4 | 85 | 35 | 500 | 120 | 2621.0 | 203.8 |
| 5 | 65 | 30 | 600 | 120 | 2599.2 | 190.6 |
| 6 | 85 | 30 | 600 | 120 | 2623.6 | 209.2 |
| 7 | 65 | 35 | 600 | 120 | 2604.9 | 196.8 |
| 8 | 85 | 35 | 600 | 120 | 2639.3 | 220.4 |
| 9 | 65 | 30 | 500 | 150 | 2522.4 | 160.3 |
| 10 | 85 | 30 | 500 | 150 | 2518.8 | 171.8 |
| 11 | 65 | 35 | 500 | 150 | 2532.0 | 164.8 |
| 12 | 85 | 35 | 500 | 150 | 2539.1 | 181.7 |
| 13 | 65 | 30 | 600 | 150 | 2553.5 | 178.1 |
| 14 | 85 | 30 | 600 | 150 | 2558.6 | 193.1 |
| 15 | 65 | 35 | 600 | 150 | 2565.2 | 182.0 |
| 16 | 85 | 35 | 600 | 150 | 2576.1 | 203.8 |
| 17 | 85 | 30 | 350 | 50 | 2875.2 | 181.6 |
| 18 | 85 | 30 | 350 | 75 | 2661.3 | 170.1 |
| 19 | 85 | 30 | 350 | 100 | 2749.2 | 149.2 |
| 20 | 85 | 30 | 350 | 125 | 2468.3 | 131.5 |
| 21 | 85 | 30 | 350 | 150 | 2370.9 | 116.0 |
| 22 | 65 | 25 | 600 | 50 | 2882.8 | 198.2 |
| 23 | 65 | 25 | 600 | 75 | 2733.9 | 188.8 |

| 24 | 65 | 25 |  | 100 | 2715.7 | 177.7 |
| 25 | 65 | 25 |  | 125 | 2590.6 | 167.1 |
| 26 | 65 | 25 |  | 150 | 2537.3 | 155.1 |
| AL-driven Bayesian optimisation cycle ||||||||
| 27 | 86 | 35 | 541 | 68 | 2985.3 | 242.4 |
| 28 | 54 | 25 | 365 | 50 | 2805.5 | 177.7 |
| 29 | 70 | 28 | 457 | 62 | 2853.2 | 204.9 |
| 30 | 45 | 27 | 539 | 50 | 2935.8 | 188.8 |
| 31 | 95 | 37 | 300 | 60 | 2947.4 | 208.6 |
| 32 | 95 | 36 | 622 | 133 | 2691.1 | 217.6 |

*Measured using an in-line Tecnar Accuraspray 4.0 particle diagnostics

## 3.2 Machine learning

### 3.2.1 ML Model prediction accuracy, data fitting, and stability

The results of the RF and GP models performing *k*-fold cross-validation on the initial dataset for particle temperature and velocities are listed in Tables 3 and 4, respectively. The accuracy of the training temperature and velocity dataset varies greatly depending on the model used, whereas the testing prediction was more accurate with the GP model. For the testing data, the GP performed the best across all metrics for both temperature and velocity. The RMSE, MAE, and $R^2$ of the GP model for predicting temperatures were 46.4, 37.5, and 0.88, respectively. Whereas for particle velocities, the RMSE, MAE, and $R^2$ of 18.2, 14.7, and 0.88, respectively, were reported.

Table 3: Accuracy results of the machine learning models when predicting temperatures.

| Model | Attributes | Training dataset ||| Testing dataset |||
|---|---|---|---|---|---|---|---|
| | | RMSE (°C) | MAE (°C) | $R^2$ | RMSE (°C) | MAE (°C) | $R^2$ |
| RF | Mean | 21.3 | 14.0 | 0.98 | 44.0 | 36.2 | 0.86 |
| | SD | 1.2 | 0.7 | | 35.4 | 29.1 | |
| GP | Mean | 40.9 | 29.2 | 0.93 | 46.4 | 37.5 | 0.88 |
| | SD | 1.9 | 1.2 | | 27.9 | 19.9 | |

Table 4: Accuracy results of the machine learning models when predicting velocities.

| Model | Attributes | Training dataset ||| Testing dataset |||
|---|---|---|---|---|---|---|---|
| | | RMSE (m/s) | MAE (m/s) | $R^2$ | RMSE (m/s) | MAE (m/s) | $R^2$ |
| RF | Mean | 8.7 | 6.1 | 0.98 | 20.0 | 16.4 | 0.86 |
| | SD | 0.7 | 0.3 | | 12.4 | 8.2 | |
| GP | Mean | 18.0 | 12.3 | 0.93 | 18.2 | 14.7 | 0.88 |
| | SD | 1.5 | 0.7 | | 12.1 | 8.2 | |

Ranking the trained models is challenging and the results could also vary slightly depending on the statistical metrics employed by the models. The RF model underperformed in 4 out of 6 metrics, with higher RMSE and MAE and a lower $R^2$ as compared to that of the GP model predictions in most cases.

Typically, the accuracy of the training data will always be higher than that of the testing data; however, a large difference in prediction accuracy may indicate poor stability, thus, directly impacting the reliability of the model (Li et al., 2020). Furthermore, frequent higher values of testing errors compared to the training errors may also indicate the model is overfitted. The overfitting of data could be detected by evaluating the difference between the training and testing errors of multiple *k*-fold iterations. Based on the data tabulated in Tables 3 and 4, the difference in training and testing error is much larger for the RF model as compared to the GP model, shown in Figure 4, which may indicate an overfitted model. For example, the relative change in data fitting ($R^2$) of training and testing data is three-fold higher in the RF model as compared to the GP model. The GP model seems to be more stable, i.e. the relative change between the training and testing accuracies, as compared to the RF model. In thermal spray, a synergetic change of ± 50 °C particle temperature and ± 20 m/s particle velocity is adequate to induce a change in deposited microstructure or associated phase changes; thus, there is a need to further improve the prediction accuracy in the GP model. As indicated in the materials and methods section, the prediction accuracy of the model is highly reliant on the informativeness of the datatset used and maximum uncertainty in the given model. Thus, it is crucial to identify informative search spaces within the given dataset and guide the sampling in the suggested search space to maximise uncertainty reduction and increase prediction accuracy.

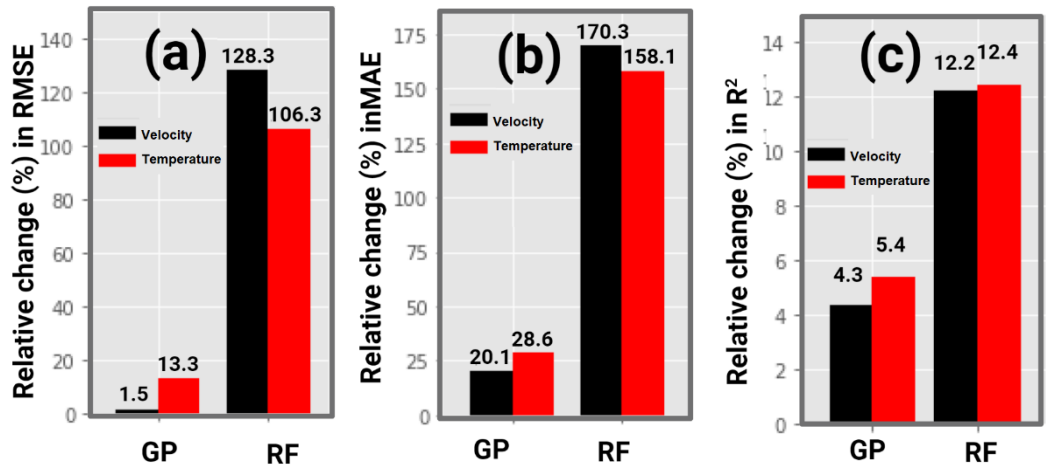

Figure 4: The relative change (%) between the training and testing accuracies with respect to (a) RMSE, (b) MAE, and (c) $R^2$ for all studied models.

### 3.3 Active learning
#### 3.3.1 Guided search

At the start of the Bayesian optimisation cycle, the training set consisted of spray trails 1-26 from the database tabulated in Table 2. To illustrate how the search was guided by Bayesian optimisation, Figures 5 and 6 are used. Figure 5 shows two scatter plots of the original experimental dataset (spray trails 1-26), where Figures 5a and 5b plot the particle temperatures and velocities at the given SODs, respectively.

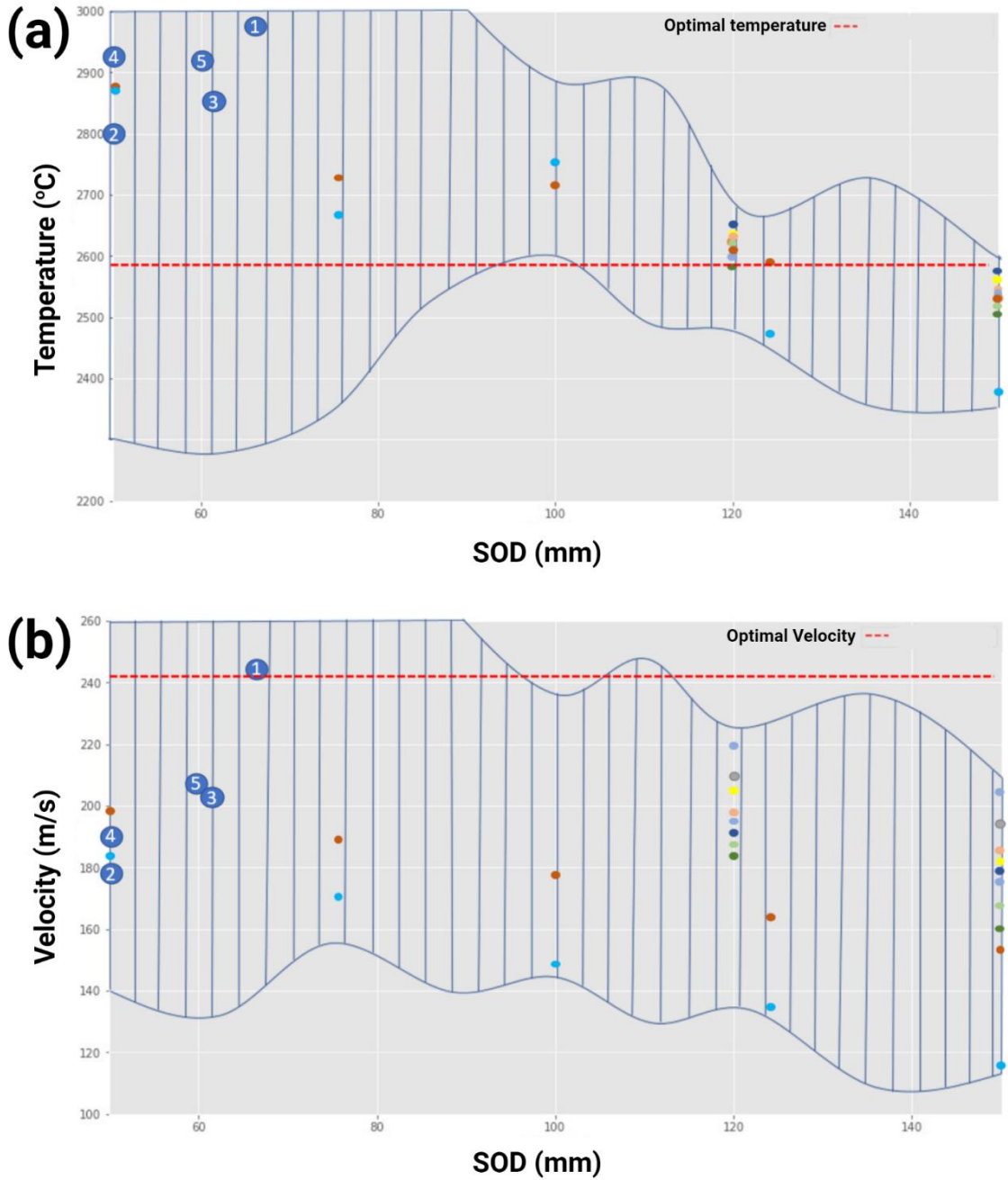

Figure 5: The scatter plots of (a) particle temperatures and (b) velocities based on probability distribution area for the initial database (1-26 spray trails).

The colour of a trail denotes specific fixed combinations of processing parameters, excluding SODs. The red dashed line represents the optimal temperature or velocity, while the dashed area approximates a probability distribution over possible functions that pass through the data points. The area, described by the GP model, illustrates the uncertainty of sampling in the dashed region. For example, in the initial experimental dataset, more experiments were conducted in the 120-150 mm SODs range. As a result, the uncertainty in the region is lower, which, by extension, leads to the hatched area being narrow when close to the data points. If a desired combination of optimal in-flight characteristics is desired at the SODs between 110-140 mm, the model will likely generalise well for the desired particle temperature; however, the same combination of processing parameters results in a reduced particle velocity than the desired velocity as the model is unable to generalise the uncertainty for a given point which is outside the region of estimated uncertainty.

In order to define a reference point where the expected improvements are desired, optimal in-flight temperature and velocity characteristics were assumed. The assumption of optimal temperature and velocity were based on the spray trails, i.e. the degree of in-flight silicone particles, and the resultant deposited microstructure, which are not discussed in this study. The choice of optimal data point also reflects the diversified framework for the optimisation, as the optimal point for particle temperatures lies firmly near the experimental dataset, which may indicate less uncertainty for the model optimisation. In contrast, the optimal particle velocity is set away from the existing experimental dataset, which is used to ascertain the overall uncertainty reduction after the optimisation.

In the SODs region from 50 to 80, the uncertainty is relatively high due to a lack of sufficient data points. As a result, the area is favoured by the acquisition function since the potential for improvement is large. The first few iterations are therefore expected to fall within this region. Based on this assumption, 5 different spray trails were conducted (spray trails 27-31) in this uncertainty region as guided by the Bayesian optimisation to reduce the maximum uncertainty in the initial database. The positioning of new spray trials in the uncertainty region is shown in Figure 5, where the numbers denote the order in which they were sampled. As these are 3 processing parameters in addition to SODs, it takes the Bayesian optimisation 5 iterations to adequately explore the shorter SODs region.

In Figure 6, the hatched uncertainty area has been updated based on the 5 samples acquired experimentally after the Bayesian optimisation cycle. After an attempt to explore the shorter SODs region of the database, it is reasonable to assume that the model may not suggest further combinations of temperature and velocity from this region. Based on the information illustrated in Figure 5, the two most likely locations to be sampled by the Bayesian optimisation were in the SODs range of 110 and 135 mm and the Bayesian optimisation iterated to sample at a SOD of 133 mm. The processing parameters suggested for spray trail 32 (point 6 in Figure 6) generated a plasma plume of 2691.1 °C and particle velocities of 217.6 m/s, which is the second-best experimental trail overall in the updated dataset. Since the Bayesian optimisation was sampling as expected for the first 6 iterations, it is reasonable to assume that it will work well for the next iterations and that further improvement would likely be made for the in-flight characteristics.

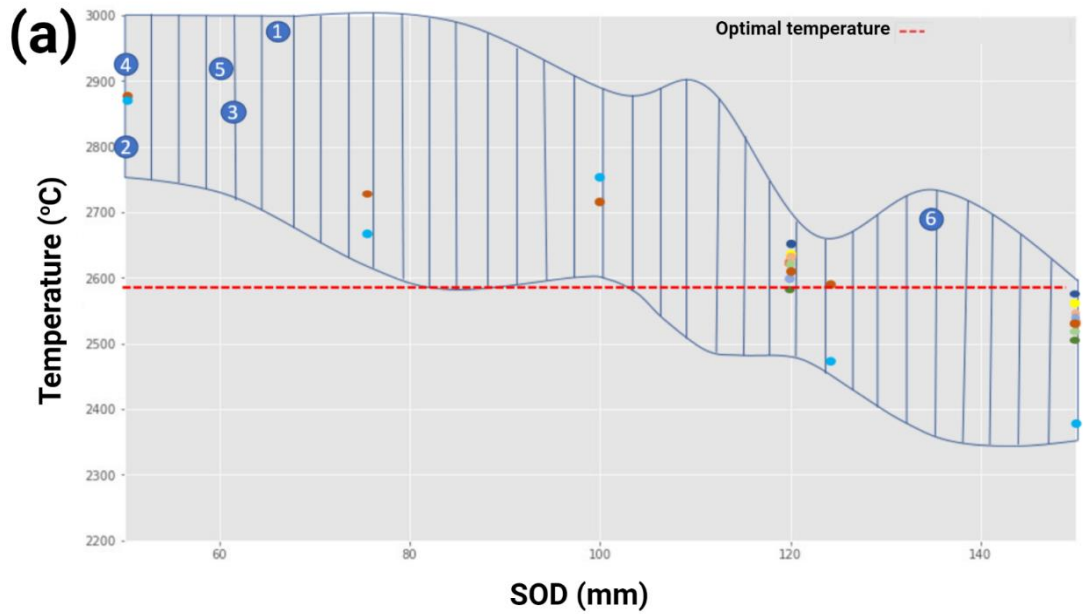

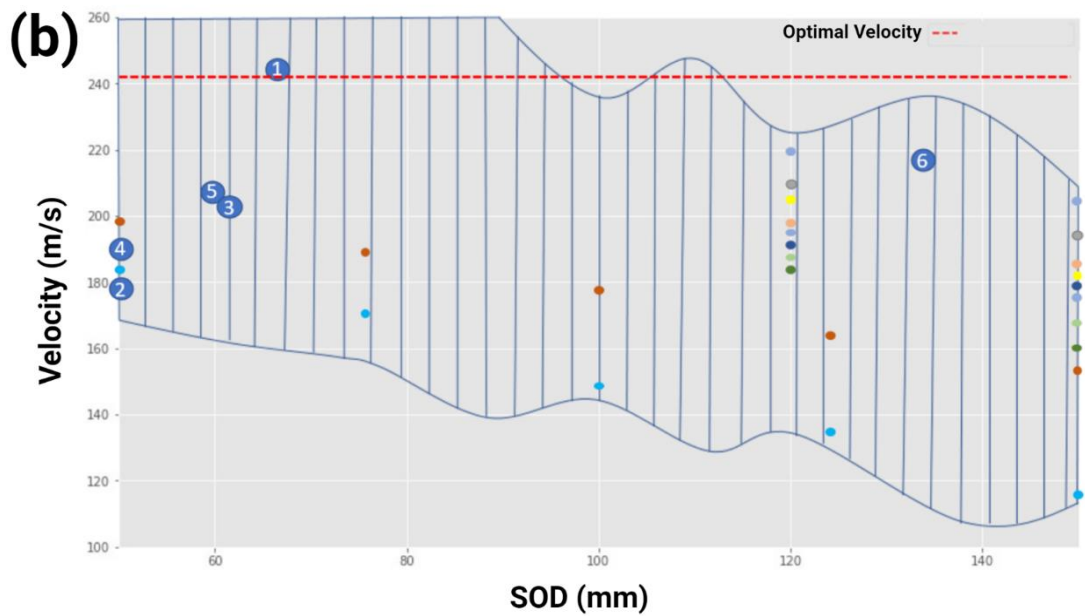

Figure 6: The updated scatter plots of (a) particle temperatures and (b) velocities based on probability distribution area for the Bayesian optimised database (1-31 spray trails).

### 3.3.2  Maximum uncertainty reduction

Based on the iterations carried out in section 3.3.1, to reach optimal processing parameters for the desired in-flight characteristics, it can be determined if the samples obtained from the AL were more informative than the samples in the original training set. How informative a sample is can be measured by looking at the reduction in maximum uncertainty after adding a new data point to the old dataset, which in this case, is the iterations carried out after the Bayesian optimisation. Table 5 shows the average reduction in maximum uncertainty either by adding an AL sample or a sample from the original database as the $i^{th}$ sample during the training of a GP model.

Table 5: Average reduction in maximum uncertainty in the dataset with respect to the positions

| Position | Original | Active Learning | % change* |
|---|---|---|---|
| 1 | 0.049 | 0.061 | 19.7 |
| 2 | 0.040 | 0.057 | 29.8 |
| 3 | 0.034 | 0.054 | 37.0 |
| 4 | 0.029 | 0.051 | 43.1 |
| 5 | 0.025 | 0.048 | 47.9 |
| 6 | 0.022 | 0.044 | 50.0 |
| 7 | 0.020 | 0.040 | 50.0 |
| 8 | 0.018 | 0.037 | 51.4 |
| 9 | 0.017 | 0.034 | 50.0 |
| 10 | 0.016 | 0.032 | 50.0 |
| 11 | 0.016 | 0.030 | 46.7 |
| 12 | 0.015 | 0.029 | 48.3 |
| 13 | 0.014 | 0.027 | 48.1 |
| 14 | 0.013 | 0.025 | 48.0 |
| 15 | 0.012 | 0.023 | 47.8 |
| 16 | 0.011 | 0.022 | 50.0 |
| 17 | 0.010 | 0.021 | 52.4 |
| 18 | 0.010 | 0.019 | 47.4 |
| 19 | 0.009 | 0.018 | 50.0 |
| 20 | 0.008 | 0.017 | 52.9 |
| 21 | 0.008 | 0.016 | 50.0 |
| 22 | 0.008 | 0.014 | 42.9 |
| 23 | 0.008 | 0.013 | 38.5 |
| 24 | 0.007 | 0.013 | 46.2 |
| 25 | 0.007 | 0.013 | 46.2 |
| 26 | 0.007 | 0.012 | 41.7 |
| 27 | 0.007 | 0.012 | 41.7 |
| 28 | 0.007 | 0.011 | 36.4 |
| 29 | 0.006 | 0.011 | 45.5 |
| 30 | 0.005 | 0.010 | 50.0 |
| 31 | 0.004 | 0.011 | 63.6 |

*The percentage change in uncertainty values as compared to original and AL values

The process of obtaining the uncertainty results consisted of several sequential steps. (i) Categorising the samples as either from the original dataset or the AL-derived dataset. (ii) Training the GP model on a single sample and generalising predictions at regular intervals over the entire bound defined search space to obtain the maximum uncertainty/standard deviation. (iii) Selecting a random unused sample (from the database), finding the maximum uncertainty in the same manner as (ii), and calculating the difference between the current maximum uncertainty and the maximum uncertainty before the current sample was added. Thus, assigning a different value relevant to the sample category and the position at which it was added. (iv) Step (iii) was repeated until all samples in the dataset are used. (v) Steps (ii-iv) were repeated $N$ number of times (30000 in this study). Lastly, (vi) all resultant values were summed, averaged, and grouped by category and the position step it was introduced. The resultant uncertainty with respect to the step position is illustrated in Figure 7. The statistical overview of the average reduction in maximum certainty is listed in Table 6. The ratio was calculated by dividing the AL uncertainty by the original uncertainty.

Table 6: The overall reduction in maximum certainty

| Uncertainties | Mean | Min | Max | SD |
| --- | --- | --- | --- | --- |
| Original | 0.015 | 0.004 | 0.049 | 0.011 |
| Active learning | 0.027 | 0.010 | 0.061 | 0.015 |
| Ratio | 1.8 | 2.5 | 1.2 | 1.4 |

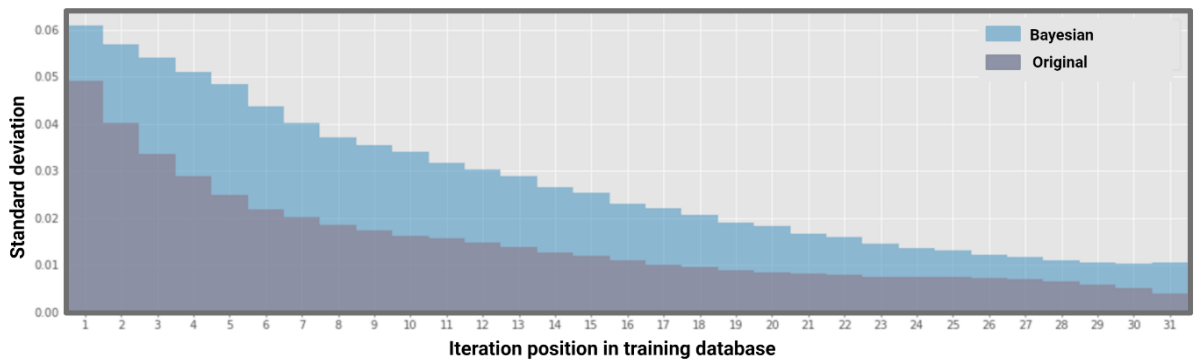

Figure 7: Reduction in maximum uncertainty as a function of step positions

The incremental step values indicated in Figure 7 shows that the standard deviation of maximum uncertainty in the database decreased with an increase in the number of spray trails. On average, the samples obtained from the Bayesian optimisation were always more informative than the samples from the original dataset, regardless of when they are introduced in training. The AL samples were on average 94.3% more informative. The smallest difference can be found at Position 1, where the AL samples are only 23.9% more informative. This is likely due to the training data dataset being relatively small at this step. The effectiveness of using AL appears to increase over time since the samples were on average 59.3% more informative for the first 5 positions and 96.0% more informative for the last 5 positions. The samples appear to have an inversely proportional relationship between how informative the sample is and at what position it was introduced, or in other words, the informativeness of a sample is a function of the iterative position it was introduced.

### 3.3.3 Contrived test validation

To check the validity of the Bayesian optimisation implementation, a custom scenario was contrived. A simple approximation, using the Euclidean distance to a specific point, was used to measure the coating quality. For this purpose, the point (2588 °C, 242 m/s), which was also the optimal in-flight particle temperature and velocity for guided uncertainty reduction, was used as the contrived test point for the distance calculations, shown in Figure 8 (the red cross). To account for any difference in units between temperature and velocity, the values were normalised before the distance was calculated. If the Bayesian optimisation adequately reduces the maximum uncertainty around the test point, then the optimisation can be adapted to accurately predict the spraying conditions for the desirable in-flight characteristics, as long as the surrogate model generalises well and the acquisition function remains unchanged.

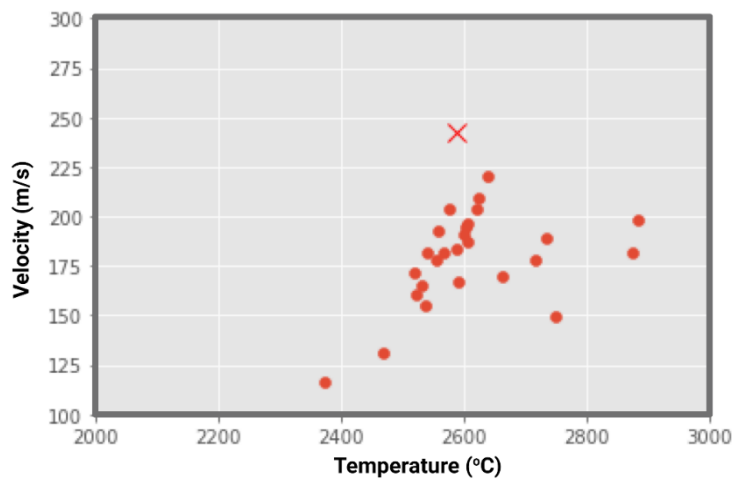

Figure 8: Chosen optimal point and initial training data

A total of ten iterations were carried out to check the validity of the Bayesian optimisation for the test point, shown in Figure 9. The initial dataset (spray trails 1-26) was used as the starting point for the first iteration of the Bayesian optimisation. As the iterations continue, the acquired samples, which are the points in each iteration with the highest EI, can be categorised as either being explorative or exploitative. Iterations 1, 2, 5, 6, 7, and 9 were mainly exploitatory, while iterations 3 and 8 were mainly exploratory. Iteration 4 seems to be a mix-match of both explorative and exploitative functions to derive the highest EI. While iteration 10 did not produce any good points to sample, thus the optimisation was terminated at this iteration.

For iteration 3, the Bayesian optimisation favoured the sample point (2000, 300) since the uncertainty is high in that region due to a lack of data points and reduced sampling in the region surrounding the optimal solution as sufficient points were sampled around the test point in iteration 2. As the likelihood of potential sampling in the regions, where the uncertainty is high due to a lack of data points, decreases and the overall EI potential increases. This allows the optimisation to favour uncertainty reductions around the test point, as shown in iterations 4-7 and 9. In the initial dataset (spray trails 1-26), the best data point nearest to the test point was spray trail 8 (2639, 220). The best-predicted data point during the Bayesian optimisation cycle after the 10 iterations carried out in this section was (2589, 249). The spray trail 8 had a Euclidean distance of 0.190, after normalising the temperature and velocity, while the predicted data point indicated a distance of 0.055, which is a three-fold increase in prediction accuracy for desired in-flight characteristics. The expected improvement after the Bayesian optimisation cycle could drastically reduce the need for explorative experimentation and can predict the best possible conditions around the desirable in-flight characteristics with the least uncertainty in the given database.

After the uncertainty reduction carried out using the AL-driven Bayesian optimisation, the RMSE and $R^2$ of the testing dataset were calculated for 1-32 spray trails and are listed in Table 7. An impressive 95.0% improvement (error reduction) of RMSE and an $R^2$ increase of 12.5% was reported on the predicted in-flight particle velocity after the AL-driven optimisation. For the in-flight particle temperature predictions, a 10.8% RMSE reduction and an $R^2$ increase of 4.5% were reported. A higher error reduction and data fitting with the in-flight velocity predictions were expected as there was a disparity in expected improvements with respect to the optimal temperature and velocity. In other words, the optimal temperature was well within the range of sampled temperature points, whereas the optimal velocity was far off the sampled velocity points, shown in Figures 5 and 6. Thus, a higher improvement was expected from the in-flight velocity predictions compared to the in-flight temperature predictions.

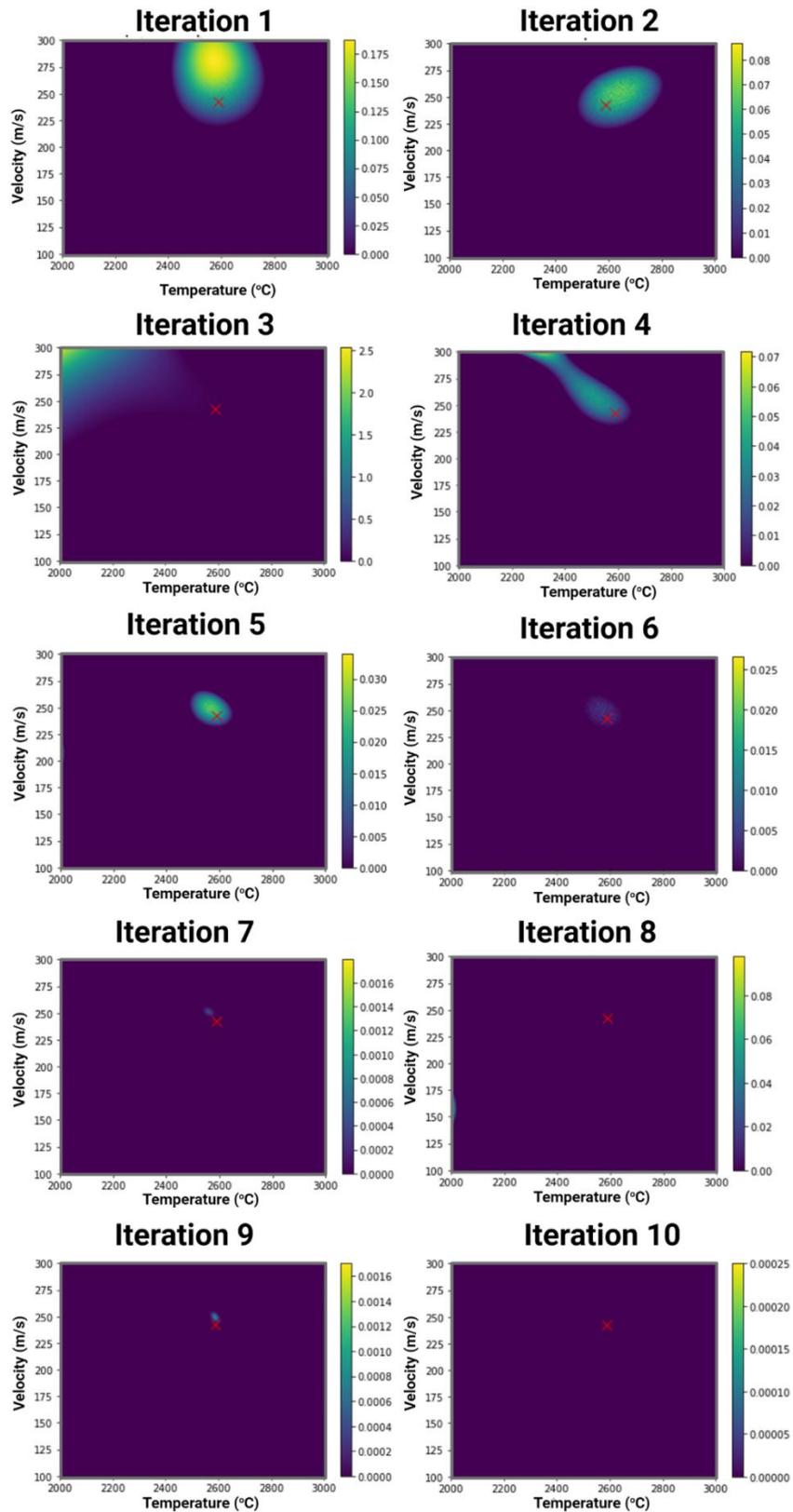

Figure 9: Bayesian Optimisation Iterations to reduce maximum uncertainty and improve EI around the contrived test point. Iterations 1, 2, 5, 6, 7, and 9 were mainly exploitatory, while iterations 3 and 8 were mainly exploratory.

Table 7: Accuracy results (testing dataset) of the GP model after the AL-driven Bayesian optimisation

| Particle characteristics | Attribute | Initial (1-26 spray trails) | AL-driven (27-32 spray trails) | Improvement |
|---|---|---|---|---|
| Temperature | RMSE (°C) | 46.4 | 41.4 | 10.8% |
| | $R^2$ | 0.88 | 0.92 | 4.5% |
| Velocity | RMSE (m/s) | 18.2 | 0.9 | 95.0% |
| | $R^2$ | 0.88 | 0.99 | 12.5% |

## 4. Discussion

In a process like thermal spraying, which has multiple chemical and thermodynamic stages (layers) and involves numerous process parameters (nodes) that are interlinked with the stages, it may seem ANNs is the best available choice to construct a neural network mimicking these complex relationships (Paturi et al., 2021). Despite the ANNs ability to learn from past data and use it to better predict the response variables, the key difference between ANNs and the statistical methods, like RF and GP models, is how the non-linear data is classified. ANNs tend to minimise the empirical risk learnt during the model training and intend to converge to local minima, which could overfit the model. Whereas the RF and GP models tend to have better generalisation capability as it intends to find a global solution, taking into consideration of the model complexities and underlying mechanisms, during the model training (Ren, 2012). In other words, ANNs inherently rely on activation functions within the multi-layer connections to deal with non-linear sampling, whereas the RF and GP models utilise a kernel function to separate the non-linear problems into linear sampling, thus, opting for a customised response unique to the non-linear problem. This makes the RF and GP models more accurate to a real-world problem with a higher degree of complexity than observed during training, and ANNs may not generalise well due to a fixed size of the neural layers that may not include a relationship with previously unseen data (Bisgin et al., 2018). In this study, two statistical ML models were used to evaluate the prediction accuracy for the given dataset and to further incorporate the best model for AL via Bayesian optimisation. Typically for a regression model, an $R^2$ value of >0.90 is considered promising for a given database and domain. In this study, the testing accuracies for the RF and GP model were below 0.90. It is imperative to mention that the applicability of an ML model largely depends on the quality or the linearity of the available database and the function or the desired combinations to generalise a prediction. Overall, the GP model performed better, in terms of prediction accuracy, data fitting, and model stability, as compared to the RF model and the GP model was used for further AL-based optimisation.

The Bayesian optimisation consists of two main components: a surrogate model and an acquisition function carried out to reduce: (I) maximum uncertainty in the database by hyperparameter tuning (further spray trails 26-32) and (II) uncertainty reduction around a contrived test point. The acquisition function, EI, was used to strike balance trade-offs between exploration and exploitation, by favouring high surrogate variance for exploration (part I) and high surrogate mean samples for exploitation (part II) (Brochu et al., 2010). Both strategies employed in this study have significance and were interlinked for AL-driven thermal spraying optimisation. In part I, an iteration of the sequential Bayesian optimisation process consists of: (i) training the surrogate model (GP in this case) on all available data (spray trails 1-26), (ii) exploration of variable combinations that maximise the acquisition function (hyperparameter tuning to maximise exploitation), (iii) run selected samples (spray trails 26-32), and (iv) update the dataset to reduce overall maximum uncertainty, thus improving the predictive accuracy in a given probability distribution of the data points. The cycle is then repeated until some termination criteria are met. This type of optimisation indicates if the prediction of the desired in-flight characteristics is possible in the given database, thus, reducing the need to carry on expensive and time-intensive thermal spraying trails that may not be physically possible or attainable based on unseen physical mechanisms and/or equipment limitations. Part II adapts a narrower approach (as compared to part I, which explores the entire database and reduces maximum uncertainty) where an improvement is expected around a desired test point. Thus, the surrogate model samples a limited region where improvement is expected and comes with the best available outcome with the least uncertainty using the Bayesian optimisation iterations. In this study, the Euclidean distance calculation was employed to calculate the distance from the predicted optimised in-flight characteristics to the optimal contrived test point. A function for estimating coating quality (numeric value) based on processing parameters would be ideal for maximisation, but it overcomplicates the existing optimisation carried out in this study. Thus, based on the Euclidean distance calculation, a coating could be classified as either "good" or bad" depending on the coating quality thresholds. In this study, the contrived test point optimisation indicated a three-fold increase in prediction accuracy as compared to the experimental dataset. When deciding if the prediction accuracy was adequate or qualified for the good/desirable in-flight particle characteristics, thresholds can be set, as indicated earlier. If the generalised outcomes fall outside of the desired thresholds, then part I optimisation can be carried out to improve the unexplored regions and decrease the current maximum uncertainty within the database. When predicting the distance to the test point, the temperatures and velocities were used to predict the coating quality instead of processing parameters. The choice has three primary benefits: (i) uses the available database to predict without further experiments, (ii) if the Bayesian optimisation gives satisfactory results for a given scenario, then the model could be adapted for other experiments, as long as the surrogate model generalises well, and (iii) the capability of the process to be visualised, with only two inputs (three if EI is counted together with particle temperature and velocity) to predict the distance.

## 5. Conclusions and outlook

In this paper, an Active Learning (AL) based approach via the Bayesian Optimisation framework was utilised to reduce the maximum uncertainty in the given database and around the contrived test scenario. Out of two Machine Learning (ML) models used (Randon Forest RF and Gaussian Process GP), the GP model was found to be suitable, in terms of prediction accuracy, data fitting, and stability, as a surrogate for the AL Optimisation.

The Bayesian Optimisation was used to identify the search spaces within the initial database (spray trails 1-26) and to guide sampling iterations (spray conditions) to reduce the maximum uncertainty. A total of six guided further spray trails (further spray trails 26-32) were carried out. On average, an impressive 52.9% improvement (error reduction) of RMSE and an $R^2$ increase of 8.5% were reported on the predicted in-flight particle velocities and temperatures after the AL-driven optimisation.

The Bayesian Optimisation was also used to reduce local uncertainty around a contrived test point to predict the best possible characteristics around the optimal in-flight particle characteristics. A Euclidean distance of 0.190, after normalising the temperature and velocity, was noted in the initial database to the test point. At the same time, the predicted data point indicated a distance of 0.055, which is a three-fold increase in prediction accuracy.

The findings in this paper indicate that the ML-driven optimisation in thermal spraying could be rapidly incorporated to solve intricate and highly nonlinear processes without necessarily utilising the underlying physical mechanisms and reducing the need to undertake resource-intensive experimentations. These AL-guided experimental validations could open ways for rapid composition development and accelerate the development of advanced coatings with minimal optimisation. Furthermore, it may also serve as the first step towards fully digital thermal spraying environments.

**Declaration of Competing Interest**

The authors declare that they have no known competing financial interests or personal relationships that could have appeared to influence the work reported in this paper.

**Acknowledgements**

This work was supported by the Engineering and Physical Sciences Research Council (EPSRC) (grant number EP/V010093/1). The authors also acknowledge the use of facilities at the Nanoscale and Microscale Research Centre of the University of Nottingham, supported by the Engineering and Physical Sciences Research Council (grant number EP/L022494/1). Some of the figures used in this study were created under an open-access license using biorender.com.

**Data availability**

The raw/processed data required to reproduce these findings cannot be shared at this time as the data also forms part of an ongoing study.